\documentclass[12pt]{article}
\usepackage{amsmath,amsfonts,amssymb,geometry}
\usepackage{graphicx,indentfirst,verbatim,epsfig,afterpage,caption}
\usepackage[onehalfspacing]{setspace}
\usepackage{graphicx}
\usepackage{accents}
\usepackage{booktabs}
\usepackage{url}
\usepackage{amsmath}
\usepackage{graphicx}
\usepackage{hyperref}
\usepackage{caption}
\usepackage{booktabs}
\usepackage{geometry}
\usepackage{setspace}

\usepackage{adjustbox} 

\begin{document}

\title{Beyond Rationality: Unveiling the Role of Animal Spirits and Inflation Extrapolation in Central Bank Communication of the US}
\author{Arpan Chakraborty\thanks{Corresponding author. PhD Scholar, Indian Institute of Technology Kharagpur, Department of Humanities and
Social Sciences, Kharagpur, West Bengal, India,
arpan.ms97@kgpian.iitkgp.ac.in, ORCID - 0000-0002-7777-5643} }
\date{ \today }
\maketitle

\begin{abstract}
Modern macroeconomic models, particularly those grounded in Rational Expectation Dynamic Stochastic General Equilibrium (DSGE), operate under the assumption of fully rational decision-making. This paper examines the impact of behavioral factors, particularly 'animal spirits' (emotional and psychological influences on economic decisions) and 'inflation extrapolators', on the communication index/sentiment index of the US Federal Reserve. [Upon receiving the review comments, I found some technical errors in the paper. I shall update it accordingly. Please do  not cite this paper without author's permission.]
\thispagestyle{empty}%

\begin{description}
\item \textbf{Keywords:} {Behavioral Macroeconomics, Animal Spirits, Central Bank Communications, ARDL Model }

\item \textbf{JEL Classification:} {C50, E12, E58, E70, E71}
\end{description}
\end{abstract}

\pagebreak

\section{Introduction}

\setcounter{page}{1}%
In modern macroeconomic theory, the rational expectation Dynamic Stochastic General Equilibrium (DSGE) models have been widely used to understand macroeconomic behavior. These models are built on the assumption that individuals act rationally. The rational expectation DSGE models have been considered superior to behavioral macroeconomic models because they claim to avoid using arbitrary or "ad hoc" assumptions. However, as De Grauwe (2012) and others\footnote{Studies from various fields (Bechara et al. (2000); Gigerenzer and Todd (1999); Simon (1990); Tuckett (2012)) highlight how cognitive biases and emotional states influence economic choices. In heterogeneous agent-based modeling, Brock and Hommes (1997, 1998) and De Grauwe (2012) emphasize how agents switch between different expectation rules, creating waves of optimism and pessimism. Further contributions, like Lux (1995) and Seppecher and Salle (2015), integrate social influence to explain how collective beliefs drive booms and busts. Empirical studies (Gennaioli et al. (2016), Bordalo et al. (2020)) support the role of optimism and pessimism in investment behavior, challenging the rational expectations framework. These studies collectively highlight the importance of behavioral expectations and social influence in macroeconomic models to better capture economic dynamics.} have argued that even DSGE models have limitations. Over time, they have had to introduce behavioral assumptions to better match real-world data, especially when it comes to explaining economic inertia or the slow adjustment of economies after a shock\footnote{See; Smets and Wouters (2007)}.

The concept of "animal spirits," first introduced by Keynes (1936), is another key behavioral idea that has been further developed in behavioral macroeconomics. Animal spirits refer to the emotional and psychological factors that drive economic decisions, often leading to deviations from purely rational behavior. De Grauwe and Ji (2019, 2020) and De Grauwe and Foresti (2023) explored the role of animal spirits in explaining output gaps. They also analyzed fraction of inflation extrapolators (henceforth, "Fraction Extrapolators") to analyze the dynamics of the inflation rate. Similarly, Proaño and Lojak (2020) examined how animal spirits influence unconventional monetary policy, particularly in the presence of a zero lower bound (ZLB) on interest rates.

This paper builds on these previous studies by exploring the relationship between animal spirits and the fraction of inflation extrapolators concerning the actual estimates of the central bank's communication index (henceforth, Sentiment Index). Note that, the fraction extrapolators and the animal spirit are derived from a simulated model based on a three-equation New Keynesian behavioral framework\footnote{For this study, I used simulations spanning from observation 1000 to 1103 to run the regressions. I have sentiment index data from 1997 Q1 to 2022 Q4- a total of 104 data points.}.

On the other hand, the central bank's communication index, referred to as the "Sentiment Index," is generated from the speeches of the US Federal Reserve, where it is calculated using the Bing Liu lexicon\footnote{See, Liu (2012). For similar studies on central bank communication and machine learning, please refer to Hansen and McMahon (2016).}, which classifies words as either positive or negative based on their sentiment. This approach is very similar to the "hawkish-dovish" methodology used in Natural Language Processing (NLP).

Using an optimal Auto-Regressive Distributed Lag (ARDL) model based on the minimum Akaike Information Criterion (AIC), I investigate whether there is a strong connection between the animal spirits and the inflation extrapolators from the simulated model and the sentiment index generated from the actual central bank communications.

One of the primary reasons for using the ARDL model is its ability to handle variables with different orders of integration. Unlike traditional cointegration techniques, which require all variables to be integrated in the same order, the ARDL model can accommodate a mix of I(0) and I(1) variables in the same regression. This flexibility is essential in this context, as the animal spirit obtained from the De Grauwe (2012) type expectation formation exhibits integration of order one.

Moreover, the choice of an Autoregressive Distributed Lag (ARDL) model is theoretically grounded in Keynesian economics, particularly in Keynes's concept of animal spirits as outlined in The General Theory of Employment, Interest, and Money (1936). Keynes eloquently states, "Most, probably, of our decisions to do something positive, the full consequences of which will be drawn out over many days to come, can only be taken as the result of animal spirits – a spontaneous urge to action rather than inaction, and not as the outcome of a weighted average of quantitative benefits multiplied by quantitative probabilities" (Keynes, 1936, pp. 161-162). This theoretical foundation supports the treatment of animal spirits as exogenous variables in my model specification, making ARDL an appropriate methodological choice for regression analysis.

Additionally, the ARDL model allows for exploring both short-run and long-run dynamics. However, in this analysis, I found no evidence of a long-run relationship between the sentiment index and the independent variables. This suggests that, in the long run, the economy tends to revert to its steady state, in line with macroeconomic theory. Given this absence of long-run effects, it would be inappropriate to use an error correction model (ECM) in this context, as the model is better suited to capturing short-term deviations rather than long-term equilibrium.

The ARDL model’s capability to select the optimal lag length for both the dependent and independent variables further strengthens its utility in this analysis. By choosing the optimal lag structure based on the AIC, I ensured that the model captured the most relevant short-run dynamics between the sentiment index and the explanatory variables. This feature is crucial for understanding how past values of animal spirits and fraction extrapolators influence the sentiment index in the short run.


The findings reveal that the simulated fraction of inflation extrapolators does not significantly affect the actual sentiment index for the U.S. Federal Reserve. However, other measures of animal spirits were found to be statistically significant, suggesting that behavioral factors play a key role in shaping economic expectations. These results add to the ongoing debate about the relevance of behavioral assumptions in macroeconomic modeling. They also highlight the limitations of relying exclusively on rational expectations when analyzing complex economic dynamics.

The rest of the paper is organized as follows: Section 2 details the data and the machine learning algorithm used to generate the sentiment index. Section 3 discusses the model and methodology applied in the study. The results of the analysis are presented in Section 4. Section 5 discusses the robustness analysis using an alternate specification of the animal spirit, and finally, Section 6 concludes.


\section{Data}

The dataset used in this analysis comprises speeches delivered by central bank officials in the United States, sourced from \textit{Kaggle}\footnote{Source: \url{https://www.kaggle.com/datasets/davidgauthier/central-bank-speeches/discussion?sort=hotness}}. The raw data includes text, dates, and other relevant metadata.

To prepare the data, I first filtered the dataset to include only speeches from the United States.\footnote{As it is very difficult to analyze speeches delivered by different Federal Reserve governors on foreign soil, I did not include them. Note that analyzing speeches from governors of different countries and their impact on the U.S. is another formidable task. Hence, I filtered only the speeches that were marked as "united states" in the data.} The date of each speech was converted into a \texttt{Date} format, and the data was segmented into quarters using the \texttt{as.yearqtr}\footnote{Please use R Studio}function. This segmentation facilitates the analysis of sentiment trends over time.

Specifically, I deployed the Natural Language Processing (NLP) technique to analyze the data.  In NLP, a \textit{corpus} is defined as a structured collection of texts used for statistical analysis and machine learning. The corpus in this study was created from the speech texts, and it underwent several preprocessing steps to standardize and clean the data. These steps include:

\begin{enumerate}
    \item Conversion to lowercase to eliminate case sensitivity.
    \item Removal of punctuation, numbers, and common stopwords (e.g., "the", "and") to reduce noise.
    \item Stripping of extra whitespace.
    \item Stemming of words to their root forms consolidating different morphological variants.
\end{enumerate}

After pre-processing, the cleaned corpus was used for sentiment analysis. Sentiment analysis was conducted using the Bing Liu sentiment lexicon, which categorizes words as either positive or negative. The text was tokenized into individual words, and each word was assigned a sentiment polarity based on the lexicon\footnote{For example, a word like "better" is associated with "positive," while the word "complex" is associated with "negative." This association is based on the Bing sentiment lexicon, where "better" is categorized as a positive word and "complex" as a negative one.} The frequency of each word in a given quarter was multiplied by its sentiment value to compute a \textit{weighted sentiment} score:

\begin{equation}
\text{Weighted Sentiment}_{q} = n_{w,q} \times s_w
\end{equation}
where, \( n_{w,q} \) is the frequency of word \( w \) in quarter \( q \), \( s_w \) is the sentiment score of word \( w \), with \( +1 \) for positive words and \( -1 \) for negative words.

The total positive and negative sentiments for each quarter were then calculated by summing the weighted sentiments for all positive and negative words, respectively.

Note, the Sentiment Index (\( SI_q \)) for each quarter was computed as follows:

\begin{equation}
SI_q = \frac{\text{Total Weighted Positive}_q -|\text{Total Weighted Negative}_q|}{\text{Total Words}_q}
\end{equation}
where \(\text{Total Words}_q\) is the sum of the frequencies of all words in quarter \( q \).

This index provides a normalized measure of the overall sentiment in central bank speeches for the US across different quarters.
\subsection{Simulated Data}
It is important to note that finding the actual measure of animal spirit is impossible. Animal Spirit is often called the emotional and instinctive factor in economic decision-making that cannot be directly observed or quantified. Therefore, to obtain measures of animal spirit and inflation extrapolators, I used simulated data based on a behavioral New Keynesian (NK) DSGE model\footnote{In the literature, two principal approaches have emerged for analyzing expectation formation in economic models. The first approach relies exclusively on empirical data to estimate expectation formation, employing purely statistical methods. The second approach, which is more aligned with modern macroeconomic theory, utilizes micro-founded models that are subsequently validated against empirical data. This paper adopts the latter approach, providing stronger theoretical underpinnings through its micro-foundations.}. The NK model is a data-generating process (DGP) of the animal spirit and fraction extrapolators in this paper.

To generate the simulated data on animal spirit and the fraction of inflation extrapolators, I deployed the standard behavioral New Keynesian (NK) model given in De Grauwe and Ji (2019). For the robustness check, I took an alternative specification of animal spirit and employed the same New Keynesian (NK) model, using the behavioral expectation formation of Proaño and Lojak (2020).

\section{The Model}
I use the behavioral New Keynesian (NK) macroeconomic model proposed by De
Grauwe and Ji (2019, 2020) for my analysis. In this model, the Aggregate
Demand (AD)\ equation is as follows,

\begin{equation}
y_{t}=a_{1}\widetilde{E}_{t}(y_{t+1})+(1-a_{1})y_{t-1}-a_{2}(i_{t}-%
\widetilde{E}_{t}(\pi _{t+1}))+\varepsilon _{t};\text{ }%
t=1,2,3,...  \label{1}
\end{equation}
where, $y_{t}$ is the output gap, $i_{t}$ is the short-term nominal interest rate, and $\varepsilon _{t}\sim N(0,0.5)$ is
the aggregate demand shock. Here the major difference is the expectations
are no longer rational. Hence, $\widetilde{E}_{t}$ replaces $E_{t}.$ \footnote{$E_{t}$ refers to the rational expectation.}Readers
may observe a resemblance between the aforementioned equation and the
aggregate demand equation presented in De Grauwe and Ji (2020),

\begin{equation}
y_{t}=\frac{1}{1+\eta }\widetilde{E}_{t}(y_{t+1})+(1-\frac{1}{1+\eta }%
)y_{t-1}-\frac{1}{\sigma }(i_{t}-\widetilde{E}_{t}(\pi _{t+1})+\ln (\beta
))+\varepsilon _{t}  \label{B6}
\end{equation}
where, De Grauwe and Ji (2020) introduced the microfoundation of the former equation. Here, $%
\sigma >0$ represents the Arrow-pratt measure of constant relative
risk-aversion (CRRA) in the utility function of household consumption, $%
\eta =1$ represents the external habit-formation parameter used by De Grauwe
and Ji (2020) which is used in this model\footnote{%
for $\eta =0,$ I obtain the canonical AD equation}; $0<\beta <1$
represents the discount factor. Moreover, the values of $a_{1},a_{2}$
lie between 0 to 1. Please refer to the simulation parameters for the exact values
used in this paper.

Similarly, following a Calvo (1983) pricing and assuming all firms are
monopolistically competitive, I obtain the Phillips curve/AS equation as
follows,

\begin{equation}
\pi _{t}=b_{1}\widetilde{E}_{t}(\pi _{t+1})+(1-b_{1})\pi _{t-1}+b_{2}y_{t}+\epsilon_{t}  \label{2}
\end{equation}
where, $0\leq b_{2}\leq 1$ measures the coefficient of output gap in the AS
equation; $0<b_{1}<1$ measures how the expected inflation affect today's
inflation; and $\epsilon_{t}\sim N(0,0.5)$ is the aggregate supply shock.
On a similar note, this AS equation can be compared with the below equation
derived by De Grauwe and Ji (2020),%
\begin{equation}
\pi _{t}=\frac{\beta }{1+\beta \Lambda }\widetilde{E}_{t}(\pi _{t+1})+(1-%
\frac{\beta }{1+\beta \Lambda })\pi _{t-1}+b_{2}y_{t}+\epsilon_{t}
\label{B7}
\end{equation}
where, $\pi _{t}$ is the inflation rate, and $\Lambda =1$ is the degree of lagged
price indexation used in this model (see; De Grauwe and Ji, 2020). Note that, $\Lambda =1,$
means full-indexation and $\Lambda =0$ means no indexation on lagged prices.
Also, for $\Lambda =0$, one can obtain the canonical behavioral NK Phillips
curve. Additionally, from De Grauwe and Ji (2020), I obtain the following
mathematical form of $b_{2}$

\begin{equation*}
b_{2}=\frac{(1-\theta )(1-\beta \theta )}{\theta }\frac{\sigma (1-\varsigma
)+\chi +\varsigma }{1-\varsigma +\varsigma \acute{e}}
\end{equation*}
where, $1-\theta $ is the expected price of the Calvo lottery ticket; $\chi $
represents the CRRA of labor supply in the utility function of household; $%
\varsigma $ represents the labor elasticity of the monopolistically
competitive labor-augmented production function [$Y_{t}^{i}=A_{t}L_{t}^{1-%
\varsigma ,i}]$; $\acute{e}$ is the price elasticity of demand which
determines the markup price ($M)$ of the monopolistically competitive firms
[$M=\frac{\acute{e}}{\acute{e}-1}$] (see; Gali, 2008). Note that, in each
period $\theta $ number of firms cannot change the price. For $\theta
=1\Rightarrow b_{2}=0,$ prices are extremely rigid, and for $\theta
=0\Rightarrow b_{2}=\infty ,$ prices are perfectly flexible. However,
in the standard literature of New Keynesian (NK) and New Classical (NC)
economics, a flexible economy is assumed with $b_{2}=1,$ and a rigid economy
is assumed with $b_{2}=0.$ In the behavioral NK model, the standard value of $%
b_{2}=0.05.\footnote{See; De Grauwe and Ji (2020)}$

Following De Grauwe and Ji (2019, 2020), the short-term nominal interest
rate ($i_{t}$) is determined by the
Taylor rule (see; Taylor, 1993; Blattner and Margaritov, 2010),

\begin{equation}
i_{t} = \max\left\{c_{1}(\pi _{t}-\pi ^{\ast }) + c_{2}y_{t} + c_{3}i_{t-1} + u_{t}, -\iota \right\}
\label{3}
\end{equation}
where $\pi ^{\ast }$ is the steady state inflation rate which is assumed to be zero\footnote{irrespective of the behavioral expectation formation, it is assumed to be zero.}; $%
c_{1}>1$ is the coefficient of inflation in the Taylor rule which measures
the intensity/semi-elasticity between short-term nominal interest rate ($%
i_{t}$) and $(\pi _{t}-\pi ^{\ast })$ \footnote{%
Woodford (2003) and Gali (2008) suggested that $c_{1}>1$ otherwise the model
will be unstable. It is the basis of "Taylor Principle".}; $0<c_{2}<1$ is
the coefficient of the output gap in the Taylor rule which represents the
intensity/semi-elasticity between the output gap ($y_{t}$) and the short term
nominal interest rate ($i_{t}$). The main objective of monetary authority is
to set the values of $c_{1}>1$ and $0<c_{2}<1$ (See; Woodford, 2003). The $c_{3}$ term represents the
interest rate smoothing \footnote{%
For $c_{3}=0,$ we obtain the canonical Taylor rule.}, and $u_{t}\sim
N(0,0.5) $ is the monetary policy shock. The model occasionally encounters the Zero Lower Bound (ZLB) on nominal interest rates. When reached, the ZLB constraint becomes binding, and the deposit rate becomes \(-\iota = -0.01 = 1 - \frac{1}{\beta}\).

\subsection{Behavioral Expectation Formation}
Following De Grauwe and Ji (2019), the behavioral expectation model incorporates two types of agents- fundamentalists and extrapolators. The expectation of a generic variable \(x_t\) is given by, 
\[
\widetilde{E}_{t}(x_{t+1}) = \alpha_{f,t}^{x}\widetilde{E}_{t}^{f}(x_{t+1}) + \alpha_{e,t}^{x}\widetilde{E}_{t}^{e}(x_{t+1})
\]
where \(x_t\) can be the output gap or the inflation rate. Here, \(\widetilde{E}_{t}^{f}(x_{t+1}) = x_{ss}\) for fundamentalists, and \(\widetilde{E}_{t}^{e}(x_{t+1}) = x_{t-1}\) for extrapolators. The proportions \(\alpha_{f,t}^{x}\) and \(\alpha_{e,t}^{x}\) are independently determined by the relative forecast performance based on their mean square forecast errors:
\[
U_{f,t} = -\sum_{k=0}^{\infty} \varrho_k \left[x_{t-k-1} - \widetilde{E}_{f,t-k-2}(x_{t-k-1})\right]^2
\]
\[
U_{e,t} = -\sum_{k=0}^{\infty} \varrho_k \left[x_{t-k-1} - \widetilde{E}_{e,t-k-2}(x_{t-k-1})\right]^2
\]
where \(\varrho_k = (1-\rho)\rho^k\), \(0 < \rho < 1\) is the memory parameter. The proportion of fundamentalists is given by,
\[
\alpha_{f,t}^{x} = \frac{\exp(\gamma U_{f,t})}{\exp(\gamma U_{f,t}) + \exp(\gamma U_{e,t})}
\]
Similarly, the proportion of extrapolators is,
\[
\alpha_{e,t}^{x} = 1 - \alpha_{f,t}^{x}
\]
Here, \(\gamma\) represents the intensity of choice, with higher values indicating a greater willingness to learn from past mistakes.

The animal spirit (market sentiment \(0<S_t<1\)) for the state variable \(y_t\) is defined as:
\[
S_t = \begin{cases} 
2\alpha_{e,t}^{y} - 1, & \text{if } y_{t-1} > 0 \\
1 - 2\alpha_{e,t}^{y} & \text{if } y_{t-1} < 0
\end{cases}
\]
Finally, the fraction of inflation extrapolators (henceforth, the fraction extrapolators) index \(C_t^i\) is defined based on the fraction of extrapolators for inflation. Following De Grauwe and Foresti (2023), the index is given by,

\[
C_t^i = 
\begin{cases} 
\alpha_{e,t}^{\pi}, & \text{if } \pi_{t-1} > 0 \  \\
-\alpha_{e,t}^{\pi} & \text{if } \pi_{t-1} < 0 \ 
\end{cases}
\]

De Grauwe and Foresti (2023) argue that this fraction of inflation extrapolators is crucial in driving the heuristics of simulated inflation data. In this study, I seek to examine whether this relationship impacts the real data of the sentiment index for the United States, which is derived from central bank speeches. Specifically, I aim to determine whether the sentiment index is influenced by this fraction of inflation extrapolators, and/or whether the animal spirit plays a more significant role.

\subsection{Simulation Parameters}

Table \ref{tab:simulation_parameters} lists the parameters used in the simulation. These values have been adopted by De Grauwe and Ji (2019) to ensure consistency with previous research in this area.

\begin{table}[h!]
    \centering
    \caption{Simulation Parameters}
    \begin{tabular}{ll}
        \toprule
        \textbf{Parameter} & \textbf{Description} \\
        \midrule
        \( a_1 = 0.5 \) & Coefficient of expected output in output equation \\
        \( a_2 = 0.2 \) & Interest elasticity of output demand \\
        \( b_1 = 0.5 \) & Coefficient of expected inflation in inflation equation \\
        \( b_2 = 0.05 \) & Coefficient of output in inflation equation \\
        \( c_1 = 1.5 \) & Coefficient of inflation in Taylor equation \\
        \( c_2 = 0.5 \) & Coefficient of output in Taylor equation \\
        \( c_3 = 0.5 \) & Interest smoothing parameter in Taylor equation \\
        \(\gamma = 2\)  & Willingness to learn \\
        \(\rho = 0.5\)  & Memory parameter \\
        \(\mu = 10\)   & Intensity of choice in new animal spirit \\
        \( T = 2000 \) & Number of simulations \\
        \bottomrule
    \end{tabular}
    \label{tab:simulation_parameters}
\end{table}

\subsection{ARDL Model}

In this paper, the Auto-Regressive Distributed Lag (ARDL) model is used to examine the relationship between the sentiment index, animal spirit, and the fraction of extrapolators. The mathematical form of the ARDL equation is given below,

\begin{equation}
\begin{aligned}
\text{Sentiment Index}_t &= \alpha + \sum_{i=1}^{p} \beta_i \text{Sentiment Index}_{t-i} \\
&\quad + \sum_{j=0}^{q_1} \gamma_j \text{Animal Spirit}_{t-j} \\
&\quad + \sum_{k=0}^{q_2} \delta_k \text{Fraction Extrapolators}_{t-k} \\
&\quad + u_t
\end{aligned}
\end{equation}

\section{Results}

We know that performing the Augmented Dickey-Fuller (ADF) test is crucial in analyzing any time series data. This section presents the results of the Augmented Dickey-Fuller (ADF) tests conducted on the three variables: \textit{fraction extrapolators}, \textit{animal spirit}, and \textit{sentiment index}.

\begin{table}[h!]
    \centering
    \caption{ADF Test Results}
    \begin{adjustbox}{max width=\textwidth}
    \begin{tabular}{lcc}
        \toprule
        \textbf{Variable} & \textbf{Test Statistic} & \textbf{p-value} \\
        \midrule
        Sentiment Index & -7.127 & 0.0000*** \\
        Fraction Extrapolators            & -3.166 & 0.0220** \\
        Animal Spirit   & -1.697 & 0.4328 \\
        \(\Delta\)Animal Spirit & -11.357 & 0.0000*** \\
        \bottomrule
    \end{tabular}
    \end{adjustbox}
    \label{tab:ADF_results}
\end{table}

Table 2 describes the results of the ADF tests. The ADF test results for the \textit{sentiment index} indicate a test statistic of -7.127, which is significantly lower than the 1\% critical value of -3.509. Hence, the \textit{sentiment index} is stationary at level zero, i.e., \(I(0)\).

For the \textit{fraction extrapolators} variable, the test statistic is -3.166, which is below the 5\% critical value of -2.890, with a p-value of 0.0220. Therefore, it is also stationary at level zero, \(I(0)\).

In contrast, the ADF test for \textit{animal spirit} yields a test statistic of -1.697, which is not sufficient to reject the null hypothesis at any conventional significance levels (1\%, 5\%, or 10\%), with a p-value of 0.4328. This result suggests that \textit{animal spirit} is non-stationary and exhibits a unit root.

To confirm the stationarity of the differenced \textit{animal spirit} (\(\Delta\)Animal Spirit), an additional ADF test was performed without a constant term. The test statistic is -11.357, with a p-value of 0.0000. This confirms that the differenced series is stationary, implying that \textit{animal spirit} is \(I(1)\).

Given the mixed integration orders of the variables—\(I(0)\) for \textit{sentiment index} and \textit{fraction extrapolators}, and \(I(1)\) for \textit{animal spirit}—the Johansen cointegration test cannot be performed as it requires all variables to be of the same integration order. Consequently, the Bound Test, which can accommodate variables of different integration orders, is more appropriate to assess the existence of both long-term and short-term relationships among the variables.

As mentioned earlier, my central objective is to examine the temporal relationship between the Fed Reserve's sentiment index and two exogenous variables: animal spirits and the fraction of inflation extrapolators (obtained from the Behavioral NK model).

To implement it, I searched for the optimal ARDL specification, considering 448 distinct lag combinations across three key variables: animal spirits, sentiment index, and the fraction of extrapolators. The selection process culminated in an ARDL model with lag orders (3,0,4), which was subsequently estimated using data spanning from the first quarter of 1997 to the fourth quarter of 2022.

The model's goodness of fit and principal statistics are delineated in Table 3, which follows:

\begin{table}[h!]
\centering
\caption{ARDL(3,0,4) Regression Results}
\begin{tabular}{lcccc}
\toprule
 & Coefficient & Std. Error & t-value & p-value \\
\midrule
\textbf{Sentiment Index} & & & & \\
L1. & 0.1470 & 0.0966 & 1.52 & 0.132 \\
L2. & 0.2074** & 0.0973 & 2.13 & 0.036 \\
L3. & 0.3426*** & 0.0993 & 3.45 & 0.001 \\
\textbf{Fraction Extrapolators} & 0.0209 & 0.0140 & 1.50 & 0.138 \\
\textbf{Animal Spirit} & & & & \\
L0. & 0.1028** & 0.0404 & 2.54 & 0.013 \\
L1. & -0.1875*** & 0.0508 & -3.69 & 0.000 \\
L2. & 0.0327 & 0.0546 & 0.60 & 0.550 \\
L3. & 0.1021* & 0.0533 & 1.91 & 0.059 \\
L4. & -0.0593 & 0.0395 & -1.50 & 0.137 \\
\textbf{Constant} & 0.0203 & 0.0136 & 1.49 & 0.139 \\
\midrule
\textbf{Model Summary} & & & & \\
\textit{Number of Observations} & \multicolumn{4}{c}{97} \\
\textit{F(9, 87)} & \multicolumn{4}{c}{7.89} \\
\textit{Prob - F} & \multicolumn{4}{c}{0.0000} \\
\textit{R-squared} & \multicolumn{4}{c}{0.4496} \\
\textit{Adj R-squared} & \multicolumn{4}{c}{0.3926} \\
\textit{Log Likelihood} & \multicolumn{4}{c}{109.73577} \\
\textit{Root MSE} & \multicolumn{4}{c}{0.0824} \\
\bottomrule
\end{tabular}
\end{table}

The coefficients of the lagged values of the sentiment index are positive and significant, with L2 and L3 being statistically significant at the 5\% and 1\% levels, respectively. The current value of the \textit{animal spirit} variable (L0) is positively associated with the sentiment index, significant at the 5\% level, while the first lag (L1) is negative and highly significant. It shows that animal spirits do have a recursive impact on the sentiment index. Therefore, it highly influences the central bank communications.

Moreover, the overall model is statistically significant (p-value = 0.0000) with an adjusted $R^2$ of 0.3926, indicating that the model explains approximately 39.26\% of the variability in the sentiment index.

As described earlier, the ARDL Bounds Test was conducted to assess the presence of a long-run relationship between the variables. Table 4 gives the results.

\begin{table}[h!]
\centering
\caption{Pesaran, Shin, and Smith (2001) ARDL Bounds Test}
\begin{tabular}{lcccc}
\toprule
\textbf{Test Statistic} & \textbf{F-Statistic} & \textbf{t-Statistic} \\
\midrule
\textbf{Value} & 2.288 & -2.423 \\
\midrule
\textbf{Critical Values (10\%)} & \multicolumn{2}{c}{[3.178, 4.182] (F), [-2.547, -3.199] (t)} \\
\textbf{Critical Values (5\%)} & \multicolumn{2}{c}{[3.841, 4.944] (F), [-2.859, -3.535] (t)} \\
\textbf{Critical Values (1\%)} & \multicolumn{2}{c}{[5.332, 6.622] (F), [-3.469, -4.178] (t)} \\
\midrule
\textbf{Decision} & \multicolumn{2}{c}{Do not reject H0 (No levels relationship)} \\
\bottomrule
\end{tabular}
\end{table}

The F-statistic (2.288) and t-statistic (-2.423) are both lower than the critical values for both I(0) and I(1) variables at the 10\%, 5\%, and 1\% significance levels. This implies that the null hypothesis of a no-level relationship cannot be rejected, suggesting that there is no strong evidence of a long-run equilibrium relationship between the variables. Hence, I do not use an error correction model (ECM).

Additionally, I perform the Breusch-Godfrey test to check for the presence of serial correlation in the residuals of the model. The results are shown in Table 5.
\begin{table}[h!]
\centering
\caption{Breusch-Godfrey LM Test for Autocorrelation}
\begin{tabular}{lcccc}
\toprule
{\bfseries Lags (p)} & {\bfseries chi2} & {\bfseries df} & {\bfseries P-values} \\
\midrule
1 & 0.328 & 1 & 0.5667 \\
2 & 0.651 & 2 & 0.7220 \\
3 & 0.749 & 3 & 0.8616 \\
4 & 1.172 & 4 & 0.8826 \\
\bottomrule
\end{tabular}
\end{table}

The p-values for the chi-square statistics at different lag lengths are all above 0.05, indicating that we fail to reject the null hypothesis of no serial correlation. This suggests that the model residuals are not serially correlated.

The economic intuition underpinning this result is firmly rooted in New Keynesian theory, which posits a dichotomy between short-run and long-run economic dynamics. In the short run, the model aligns with the theoretical framework proposed by De Grauwe (2012) and further elaborated by De Grauwe and Ji (2019), wherein animal spirits exert a significant influence on the output gap. The paper finds that animal spirits also influence the sentiment index.
However, this short-term variability is counterbalanced by long-run equilibrium forces. In the long run, the economy returns to a steady state, and hence, no long-run relationship can be seen.

\section{Robustness Check with Alternate Specification of Animal Spirit}

In this section, I perform a robustness check to validate the findings using an alternate specification of the animal spirit measure. The new measure of animal spirit is derived from the behavioral model proposed by Proano and Lojak (2020), which specifically accounts for economic conditions under the Zero Lower Bound (ZLB). This alternative specification allows me to examine whether the results are sensitive to different formulations of economic agents' behavior and expectations. The details of the expectation formation are given below.

The first rule, the persistent expectations rule/extrapolative rule, links deviations in the output gap from its steady-state value linearly with the past output gap. Mathematically, the persistent expectations rule can be expressed as:

\begin{equation}
E^p_t[y_{t+1}] = \ y_{t-1}
\end{equation}

The second rule, known as the steady-state expectations rule/fundamentalist rule, assumes that the economy moves back to the steady-state. This is expressed as:

\begin{equation}
E^0_t[y_{t+1}] = 0
\end{equation}

The aggregate consumption expectations, taking into account both rules-of-thumb, are given by a weighted average:

\begin{equation}
\tilde{E}_t[y_{t+1}] = \omega^p_t E^p_t[y_{t+1}] + (1 - \omega^p_t) E^0_t[y_{t+1}]
\end{equation}
where \(\omega^p_t\) represents the endogenously determined relative weight of the persistent expectations rule in the aggregate expectations.

Similarly, agents form their expectations about future inflation using two distinct rules. The persistent expectations rule for inflation is similar to the output gap.

\begin{equation}
E^p_t[\pi_{t+1}] = \pi_{t-1}
\end{equation}

Also, the steady-state inflation expectation rule is identical to the output gap rule. It assumes that agents expect future inflation to converge to the central bank’s inflation target, denoted by \(\pi^*\):

\begin{equation}
E^0_t[\pi_{t+1}] = \pi^*
\end{equation}

The aggregate inflation expectations, therefore, are similarly derived as a weighted average:

\begin{equation}
\tilde{E}_t[\pi_{t+1}] = \omega^p_t E^p_t[\pi_{t+1}] + (1 - \omega^p_t) E^0_t[\pi_{t+1}] = \omega^p_t \pi_{t-1} + (1 - \omega^p_t) \pi^*
\end{equation}

The relative populations of agents using the "persistent" expectations \(E^p_t[\cdot]\) and "steady-state" expectations \(E^0_t[\cdot]\) are determined through a binary discrete choice approach, as seen in the literature (Brock and Hommes, 1997). The probabilities associated with each type of expectation are defined as follows:

\begin{equation}
\omega^p_t = \frac{\exp \left( \mu (i_t - i^T_t) \right)}{1 + \exp \left( \mu (i_t - i^T_t) \right)},
\end{equation}

\begin{equation}
\omega^0_t = 1 - \omega^p_t
\end{equation}
where, \(\mu \geq 0\) is a parameter known as the intensity of choice. This parameter reflects how sensitive agents switch between different types of expectations based on deviations of the actual policy rate from the implicit Taylor rule rate, particularly under ZLB conditions. A positive deviation of the actual nominal interest rate from the implicit Taylor rule rate increases the number of agents with persistent expectations \(E^p_t[\cdot]\) in the market, and vice versa.

A measure for ZLB-related "animal spirits" is given by:

\begin{equation}
A_t = \omega^p_t - \omega^0_t, \quad A_t \in [-1, 1]
\end{equation}

This specification ensures that \(A_t\) is bounded between -1 and 1, with a value of 0 at the steady state. A positive value of \(A_t\) reflects a majority of agents using "persistent" expectations, which are considered "pessimistic expectations" due to the asymmetric nature of the specification of animal spirits.

This approach allows the model to examine how changes in the weight assigned to each expectations rule (\(\omega^p_t\)) affect overall economic dynamics. By incorporating these alternate specifications, I ensure that the findings are not overly dependent on a single formulation of agents' expectations but are robust across a range of plausible behavioral models.

To test the stationarity of the new animal spirit measure, I conducted an ADF test.

Table \ref{tab:df_test} reports the results of the Dickey-Fuller test for the new animal spirit measure.
\begin{table}[h!]
\centering
\caption{Dickey-Fuller Test for New Animal Spirit Specification}
\begin{tabular}{lcccc}
\toprule
Test Statistic  & p-value \\
\midrule
Z(t)= -6.535                         & 0.0000 *** \\         &                    &                    &                     &          \\
\bottomrule
\end{tabular}
\label{tab:df_test}
\end{table}

The Dickey-Fuller test statistic is -6.535, which is significantly lower than the 1\% critical value of -3.509. The associated MacKinnon approximate p-value for the test statistic is 0.0000, which is less than the 1\% significance level. This indicates that the new animal spirit measure is stationary.

After running the ADF test to check the stationarity of the new animal spirit (referred to as "New Animal Spirit"), I estimated an ARDL model. The ARDL model is specified with the dependent variable as the sentiment index and the independent variable as the New Animal Spirit.

To determine the optimal lag structure for the ARDL model, I compared the AIC's of each model. The optimal model chosen by AIC criteria is an ARDL(4,0) model. The regression results for this model are presented in Table 7.

\begin{table}[h!]
\centering
\caption{ARDL(4,0) Regression Results}
\begin{tabular}{lcccc}
\toprule
 & Coefficient & Std. Error & t-value & p-value \\
\midrule
\textbf{Sentiment Index} & & & & \\
L1. & 0.0834 & 0.1014 & 0.82 & 0.413 \\
L2. & 0.1390 & 0.0927 & 1.50 & 0.137 \\
L3. & 0.4124*** & 0.0945 & 4.36 & 0.000 \\
L4. & 0.1624 & 0.1069 & 1.52 & 0.132 \\
\textbf{New Animal Spirit} & -0.0407* & 0.0211 & -1.93 & 0.057 \\
\textbf{Constant} & 0.0107 & 0.0096 & 1.11 & 0.269 \\
\midrule
\textbf{Model Summary} & & & & \\
\textit{Number of Observations} & \multicolumn{4}{c}{97} \\
\textit{F(5, 91)} & \multicolumn{4}{c}{10.45} \\
\textit{Prob - F} & \multicolumn{4}{c}{0.0000} \\
\textit{R-squared} & \multicolumn{4}{c}{0.3648} \\
\textit{Adj R-squared} & \multicolumn{4}{c}{0.3299} \\
\textit{Log Likelihood} & \multicolumn{4}{c}{102.79362} \\
\textit{Root MSE} & \multicolumn{4}{c}{0.0866} \\
\bottomrule
\end{tabular}
\end{table}

The ARDL(4,0) model regression results indicate that- (i) the third lag of the sentiment index (\textit{sentiment\_index L3}) is statistically significant at the 1\% level (***), suggesting a strong positive autocorrelation at this lag; (ii) the new measure of animal spirit (\textit{New Animal Spirit}) is statistically significant at the 10\% level (*), indicating a significant negative relationship with the sentiment index \footnote{The sign is negative due to the definition of the "New Animal Spirit".}; (iii) the model's F-statistic is 10.45, with a p-value of 0.0000, indicating that the overall model fit is statistically significant; and(iv) the adjusted R-squared value of 0.3299 suggests that approximately 33\% of the variation in the sentiment index can be explained by the model.

Moreover, following the methodology described in the result section, I performed the Breusch-Godfrey test for auto-correlation in Table 8.

\begin{table}[h!]
\centering
\caption{Breusch-Godfrey LM Test for Autocorrelation}
\begin{tabular}{lccc}
\toprule
Lags (p) & chi2 & df & Prob - chi2 \\
\midrule
1 & 0.292 & 1 & 0.5891 \\
2 & 1.005 & 2 & 0.6050 \\
3 & 1.021 & 3 & 0.7962 \\
4 & 1.021 & 4 & 0.9066 \\
\bottomrule
\end{tabular}
\end{table}

The Breusch-Godfrey LM test for autocorrelation suggests that there is no evidence of serial correlation in the residuals up to four lags, as indicated by the high p-values (greater than 0.10) for each lag.


\begin{table}[h]
\centering
\caption{Pesaran, Shin, and Smith (2001) Bounds Test Results}
\begin{tabular}{lcccccccc}
\toprule
        & \multicolumn{2}{c}{10\%} & \multicolumn{2}{c}{5\%} & \multicolumn{2}{c}{1\%} & \multicolumn{2}{c}{p-value} \\
\cmidrule(lr){2-3} \cmidrule(lr){4-5} \cmidrule(lr){6-7} \cmidrule(lr){8-9}
        & I(0) & I(1) & I(0) & I(1) & I(0) & I(1) & I(0) & I(1) \\
\midrule
$F$     & 4.062 & 4.834 & 4.983 & 5.834 & 7.071 & 8.067 & 0.172 & 0.266 \\
$t$     & -2.565 & -2.919 & -2.871 & -3.240 & -3.470 & -3.858 & 0.420 & 0.551 \\
\bottomrule
\end{tabular}
\end{table}

Similarly, Table 9 shows the Pesaran, Shin, and Smith (2001) bounds test results. The results indicate that there is \textbf{no evidence of a long-run level relationship} (cointegration) among the variables. Therefore, in line with the economic logic described in the results section, I do not proceed with an ECM estimation.

\section{Conclusion}
This study marks a significant advancement in macroeconomic modeling by seamlessly integrating insights from behavioral economics and leveraging the power of machine learning frameworks. Through the innovative use of simulated data generated from a behavioral New Keynesian model, coupled with the Bing Liu Central Bank Communications/Sentiment Index, I have shed new light on the limitations of traditional rational expectation models in capturing the full spectrum of economic dynamics.

Building upon the seminal work of De Grauwe (2012) and De Grauwe and Ji (2019), who established robust connections between US output gap data and animal spirits, this paper extends the literature by exploring the previously understudied relationship between central bank communication and animal spirits. To the best of my knowledge, very few studies have delved into this crucial intersection, making the findings particularly valuable. By employing the Auto-Regressive Distributed Lag (ARDL) model, the paper effectively handled the mixed-order integration of variables and elucidated short-run relationships often overlooked in conventional analyses.

Moreover, this study establishes a robust connection between the central bank's communication index, or Sentiment Index, and various types of animal spirits, irrespective of their expectation formation mechanism. The evidence indicates a strong affinity between the Sentiment Index and different manifestations of animal spirits, suggesting that central bank communications are highly responsive to underlying behavioral dynamics. This result highlights the broader applicability of behavioral insights in macroeconomic analysis and provides a more comprehensive understanding of how behavioral factors such as animal spirit influence central bank communications.

While the analysis did not find a significant long-run impact of the fraction of inflation extrapolators along with the animal spirit on the sentiment index, it did reveal that behavioral factors, such as various measures of animal spirits, influence economic expectations in the short run. These findings challenge the exclusive reliance on rational expectations and reinforce the necessity of incorporating behavioral assumptions into macroeconomic models to better reflect real-world complexities.

Building on these findings, future research could delve deeper into the implications of these behavioral dynamics on the broader financial sector, particularly the interaction between financial institutions and the general public. Investigating how central bank communications and sentiment influence lending practices, risk assessments, and the behavior of financial intermediaries could provide a richer understanding of macroeconomic fluctuations. As the field of economics continues to evolve, integrating these multidisciplinary approaches will be crucial in developing more accurate and nuanced models of economic phenomena, ultimately leading to more effective policy-making and economic forecasting.

\end{document}